\documentclass[11pt]{article}

\usepackage[preprint]{acl} 

\usepackage{times}
\usepackage{latexsym}

\usepackage[T1]{fontenc}

\usepackage[utf8]{inputenc}

\usepackage{microtype}

\usepackage{inconsolata}

\usepackage{graphicx}

\usepackage{xspace}

\newcommand{\dataset}{\textbf{\textsc{LAKE}}\xspace}
\newcommand{\datasetasr}{\textbf{\textsc{LAKE-asr}}\xspace}

\definecolor{over}{HTML}{D97706}
\definecolor{under}{HTML}{2563EB}
\newcommand{\cswpos}[2]{\cellcolor{over!#1}#2}
\newcommand{\cswneg}[2]{\cellcolor{under!#1}#2}
\author{%
\textbf{Junyeong Park}$^{\spadesuit}$\,
\textbf{Jieun Han}$^{\spadesuit}$\,
\textbf{Yeon Su Park}$^{\spadesuit}$\,
\textbf{Youngbin Lee}$^{\heartsuit}$\\
\textbf{Suin Kim}$^{\heartsuit}$\,
\textbf{Juho Kim}$^{\spadesuit}$\,
\textbf{Alice Oh}$^{\dagger,\spadesuit}$\,
\textbf{So-Yeon Ahn}$^{\dagger,\spadesuit}$\\
$^{\spadesuit}$KAIST, $^{\heartsuit}$Elice Inc.\\
{\{\texttt{junyeong.park@kaist.ac.kr, alice.oh@kaist.edu, ahnsoyeon@kaist.ac.kr}\}}
}

\title{When Language Learners Code-Switch:\\ Korean–English Code-Switching in LLM-based Speaking Practice}

\usepackage{lipsum}
\usepackage{kotex}
\usepackage{booktabs}
\usepackage{tabularx}
\usepackage{multirow}
\usepackage{pifont}
\usepackage[table, dvipsnames]{xcolor}
\usepackage[most]{tcolorbox}
\usepackage{enumitem}
\usepackage{graphicx}
\usepackage{subcaption}
\usepackage{hyperref}

\begin{document}
\maketitle

\begin{abstract}
Code-switching (CSW), the use of multiple languages within the same interaction, has been widely studied in NLP for multilingual communication.
While CSW also occurs in language-learning contexts, the functions and pedagogical implications of learner CSW in learner--AI interactions remain underexplored.
To address this gap, we introduce \dataset, a dataset of \textbf{L}earner--\textbf{A}I dialogues with \textbf{K}orean-\textbf{E}nglish CSW. This dataset captures South Korean learners' CSW behavior during English speaking practice with LLM-based conversation partners, comprising 675 dialogues, 19.6K utterances, and 43.9 hours of learner speech.
We develop a taxonomy of learner CSW functions, finding that learners switch to Korean for diverse purposes, including signaling lexical gaps, negotiating meaning, and conveying culturally specific content.
We also observe learners reusing and adapting English expressions introduced by the LLM partner following their code-switches, suggesting opportunities for contextualized lexical support.
To support research on speech recognition for learner CSW, we release 4.9 hours of code-switched learner speech with human-verified transcripts and benchmark speech-to-text models on Automatic Speech Recognition (ASR), revealing difficulties in recognizing learner CSW.
These findings highlight learner CSW as a communicative and pedagogically meaningful resource that can inform the design of AI-based conversational language learning systems.
\footnote{The dataset will be made public upon publication.}
\end{abstract}

\section{Introduction}
\begin{figure}[t]
    \centering
    \includegraphics[width=0.9\columnwidth]{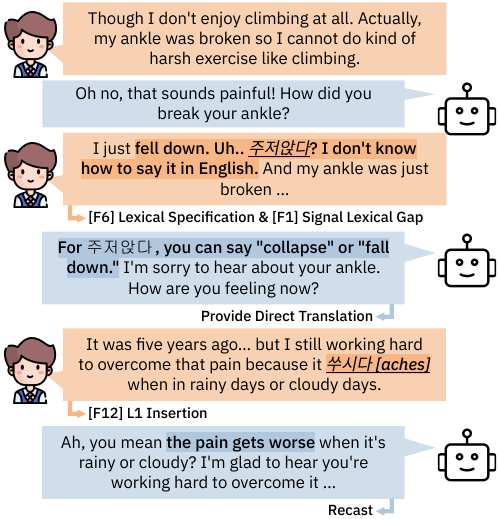}
    \caption{A South Korean student's CSW dialogue in LLM-based English speaking practice (Sample from \dataset)}
    \label{fig:figure1}
\end{figure}

Code-switching (CSW), the use of two or more languages within the same interaction, is a common practice among multilingual speakers~\cite{poplack1980csw, myers1993csw}.
Recognizing the rich social and cultural meaning of this behavior, prior NLP research has invested significant effort into curating CSW datasets~\cite{lyu2010seame, lovenia-etal-2022-ascend} and building models that can appropriately support CSW interactions~\cite{oh-etal-2026-ola, sheth-etal-2026-beyond}.

Beyond everyday multilingual communication, CSW plays an important role in \textit{language education}.
Language learners report that switching to their first language (L1) can help alleviate anxiety when speaking in the target language (L2)~\cite{sholikhah2024efl}.
Learners also switch to L1 as a strategic, compensatory move to resolve communication breakdowns in L2, potentially revealing where linguistic support is needed~\cite{moore2002code}.
Teachers can use these moments to help learners express their intended meaning in L2~\cite{ziegler2012student}.
With LLMs being increasingly used for conversational language practice~\cite{ribeiro-flucht-etal-2025-framework}, understanding when and why learners code-switch is important for informing how these systems respond to learners' communicative needs.


\newcommand{\cmark}{\textcolor{Green}{\ding{51}}}
\newcommand{\xmark}{\ding{55}}

\begin{table*}[t]
\centering

\small
\setlength{\tabcolsep}{3.5pt} 
\renewcommand{\arraystretch}{1.1} 

\begin{tabular}{
    @{}
    p{2.7cm} 
    l
    l
    l
    l
    l
    r
    r
    c
    @{}
}
\toprule
\textbf{Dataset} &
\textbf{Modality} &
\textbf{Lang.} &
\textbf{Context} &
\textbf{Speaker} &
\textbf{Interaction} &
\textbf{Uttr.} &
\textbf{Audio} &
\textbf{Func.} \\
\midrule

SEAME \newline \cite{lyu2010seame} &
Speech &
ZH-EN &
Conversation &
Bilingual &
Human--Human &
25.0K &
$\sim$30.0 h &
\xmark \\

ASCEND \newline \cite{lovenia-etal-2022-ascend} &
Speech &
ZH-EN &
Conversation &
Bilingual &
Human--Human &
12.3K &
10.6 h &
\xmark \\

TALCS \newline \cite{li2022talcs} &
Speech &
ZH-EN &
Online L2 teaching &
L2 Teacher &
Teacher--Learner &
370K &
587.0 h &
\xmark \\

RECIPE4U \newline \cite{han-etal-2024-recipe4u} &
Text &
KO-EN &
Essay revision &
L2 Learner &
Learner--AI &
4.3K &
-- &
\xmark \\

WriteAid-Data \newline \cite{myung2026scaffolding} &
Text &
KO-EN &
Essay writing &
L2 Learner &
Learner--AI &
14.9K &
-- &
\xmark \\

\rowcolor{gray!15} \dataset &
Speech &
KO-EN &
L2 speaking practice &
L2 Learner &
Learner--AI &
19.6K &
43.9 h &
\cmark \\

\bottomrule
\end{tabular}
\caption{Comparison of \dataset with existing CSW datasets. (Func. is CSW function annotation)}
\label{tab:dataset_comparison}
\end{table*}

Recent efforts are beginning to explore CSW in educational contexts.
From the teacher perspective, TALCS~\cite{li2022talcs} introduces a dataset of language teachers' CSW utterances in online classes, and \citet{liu-etal-2024-optimizing} propose a pedagogy-inspired framework to evaluate how effectively LLM tutors use CSW for pedagogical support.
From the learner perspective, RECIPE4U~\cite{han-etal-2024-recipe4u} and WriteAid-Data~\cite{myung2026scaffolding} collect learner--AI interactions in essay writing, which include written learner CSW instances.
However, a gap still remains in understanding learner CSW in \textit{spoken} interactions with AI partners, as well as how effectively current speech systems recognize such learner CSW.

To address this gap, we introduce \dataset, a dataset of \textbf{L}earner--\textbf{A}I dialogues with \textbf{K}orean-\textbf{E}nglish CSW.
This dataset captures South Korean learners' CSW behavior during English speaking practice with LLM-based conversation partners.
It contains 675 dialogues, 19.6K utterances, and 43.9 hours of learner speech, including 247 dialogues (36.6\%) containing code-switched learner utterance.
We further release 4.9 hours of learner CSW speech with human-verified transcripts to serve as a benchmark for evaluating Automatic Speech Recognition (ASR) performance on speech-to-text (STT) models.

Based on this dataset, we develop a taxonomy of learner CSW functions and manually annotate each CSW instance.
We observe that learners switch to L1 for diverse purposes, including signaling lexical gaps, managing meaning across languages, and conveying culturally specific content (Figure~\ref{fig:figure1}).
We further observe instances where learners reuse or adapt L2 expressions introduced by the LLM partner following a code-switch, demonstrating how learner CSW can create opportunities for contextualized linguistic support.
We also compare LLM-simulated learners with \dataset, finding substantial differences in CSW behavior.
Finally, our benchmark results show that current STT models struggle with learner CSW, with even the lowest Mixed Error Rate exceeding 20\%, highlighting speech recognition as an additional bottleneck in developing AI speaking partners for language learners.

In summary, our contributions are as follows:
\begin{itemize}[topsep=1pt, itemsep=1pt]
\item We introduce \dataset, a corpus of 675 learner--AI dialogues and 43.9h of learner speech, containing Korean--English learner CSW in English speaking practice with an AI partner.
\item We develop a taxonomy of learner CSW functions, and analyze how learners code-switch and subsequently reuse L2 expressions introduced by the LLM partner.
\item We release a 4.9 hour learner CSW speech dataset with human-verified transcripts and benchmark current ASR models.
\end{itemize}

\section{Related Work}
\subsection{Learner CSW in Language Education}
Learner CSW has been studied primarily in second- and foreign-language classrooms.
Switching to the first language (L1) can help learners alleviate anxiety when speaking in the target language (L2)~\cite{sholikhah2024efl} and sustain communication when they encounter lexical gaps~\cite{eldridge1996code,liebscher2005learner}.
During collaborative tasks, learners also use L1 to check their understanding and coordinate with peers, supporting task completion beyond their current L2 proficiency~\cite{swain2000task,anton1998socio}.
CSW can also serve pragmatic functions, such as marking a side comment within an L2 conversation~\cite{liebscher2005learner}.
Together, these studies show that learner CSW serves multiple functions and should not be interpreted solely as evidence of limited L2 proficiency.
More recently, RECIPE4U~\cite{han-etal-2024-recipe4u} and WriteAid-Data~\cite{myung2026scaffolding} have collected learner--AI interactions in essay writing that include learner CSW.
Our work extends this line of research to spoken learner--AI interactions, examining the functions of learner CSW during conversational language practice.

\subsection{Spontaneous CSW Speech Datasets}

CSW speech datasets have been developed primarily for the ASR task.
Prior work has collected naturally occurring CSW speech primarily from \textit{bilingual} and \textit{multilingual} speakers.
Existing corpora collect CSW from informal conversations and interviews~\citep{lyu2010seame,li-etal-2012-mandarin,hamed-etal-2020-arzen,lovenia-etal-2022-ascend}, and teachers' CSW from online language-teaching sessions~\citep{li2022talcs}.
Others collect CSW speech from online sources, such as multilingual YouTube videos~\cite{singh2024mecos, yan-etal-2026-cs}.
Our dataset complements these corpora by capturing learner CSW during L2 speaking practice and providing an ASR evaluation set for learner CSW.

\section{Dataset Construction}
This section explains how we collect (\S\ref{sec:data_participant}--\S\ref{sec:data_system}), post-process (\S\ref{sec:data_post_process}), and annotate (\S\ref{sec:data_csw_function}) the dataset, and report the final statistics in \S\ref{sec:data_final}.

\subsection{Participant Recruitment}
\label{sec:data_participant}
We recruited 40 English learners from universities in South Korea. All participants were native Korean speakers with limited exposure to English-speaking environments, including overseas experience. Their English speaking proficiency ranged from \textit{below basic} to \textit{advanced} (Detailed demographics are provided in App~\ref{app:participant_info}).

\subsection{Data Collection Procedure}
\label{sec:data_procedure}

We collected learner--AI dialogues from six weeks of English speaking practice with an LLM-based conversation partner.
To capture naturally occurring learner CSW across different contexts of shared L1 understanding, we collected dialogues with partners prompted to act as though they understood only English (\textit{English-only}, EO) or to understand both English and Korean (\textit{Korean-aware}, KA).
We initially assigned 20 participants to each setting, balancing English proficiency across settings. One participant in EO later withdrew, leaving 19 in EO and 20 in KA.
We did not instruct or encourage participants to code-switch.

\subsection{LLM-based Speaking Practice System}
\label{sec:data_system}
The system processes each learner turn through ASR, response generation, and audio synthesis.

\paragraph{ASR.}
We transcribed learner speech using \texttt{Whisper large-v2}~\cite{radford2023robust} with speech-based in-context learning~\cite{wang2024whisper}.
In the KA setting, learners could optionally press a language toggle while speaking to indicate a language switch.
When the toggle was used, we instead split the speech at the indicated switch points and transcribed the resulting segments separately using \texttt{Whisper large-v2} for English and \texttt{ReturnZero}\footnote{\url{https://developers.rtzr.ai/}} for Korean.

\paragraph{Response generation.}
We used \texttt{GPT-4o}~\cite{hurst2024gpt4o} prompted to act as a friendly, patient English conversation partner for casual discussions across topics.
In EO, the partner was instructed to act as an English-only conversation partner that can only understand English.
In KA, it was instructed to understand Korean and respond according to context: introducing an English equivalent explicitly or implicitly, asking about the Korean expression, or continuing without addressing the switch.
The KA opening greeting also stated that it understands Korean: \textit{``Hi, I'm Emily. I can speak both English and Korean. How are you?''}
(See prompts in App~\ref{app:system_response_gen}).

\paragraph{Audio synthesis.}
We synthesized partner responses using \texttt{Google Cloud Text-to-Speech}\footnote{\url{https://docs.cloud.google.com/text-to-speech/docs/}}.

\subsection{Data Post-Processing}
\label{sec:data_post_process}
We post-processed the dialogues through transcription verification, personally identifiable information (PII) anonymization, token-level language labeling, and translation. 

\paragraph{Transcription verification.}
To identify code-switched utterances that the system may have missed, we generated two additional transcripts for each learner utterance using open models: \texttt{Whisper large-v3}~\cite{radford2023robust} and \texttt{Qwen3-ASR-1.7B}~\cite{shi2026qwen3asr}.
Using regular expressions and \texttt{Gemma-4-12B-IT}~\cite{gemmateam2026gemma4}, we flagged utterances containing Korean or romanized Korean in any of the three transcripts, including the original transcript (See prompt in App~\ref{app:post_transcription_cleaning}).
A human evaluator manually verified each flagged utterance against the audio and corrected its transcript when necessary.

\paragraph{PII anonymization.}
We manually inspected dialogues to identify recurring PII categories, finding learner names, universities, cities of residence, dates, and ages. Using these categories and examples, we used \texttt{Gemma-4-12B-IT} to detect and replace PII with type-specific placeholders such as \texttt{[NAME]}, \texttt{[LOC]}. A human evaluator reviewed the entire corpus to check the replacements (See prompt in App~\ref{app:post_pii_anonymization}).

\begin{table}[thbp]
\centering
\small
\setlength{\tabcolsep}{4pt}

\begin{tabularx}{\columnwidth}{@{}l l X@{}}
\toprule
\textbf{Label} & \textbf{Lang} & \textbf{Example}\\
\midrule
\texttt{EN}   & EN    & food, restaurant\\
\texttt{EN-b} & EN    & kimchi, soju, taekwondo\\
\texttt{KO}   & KO    & 방어 (\textit{yellowtail}), 자유형 (\textit{freestyle})\\
\texttt{NE-*} & KO    & 전주 (\textit{Jeonju}), 제주 (\textit{Jeju})\\
\texttt{MIX}  & EN+KO & 나물s (\textit{namul-s}), 떡볶이s (\textit{tteokbokki-s})\\
\texttt{ANM}  & --    & \texttt{[LOC]}, \texttt{[NAME]}, \texttt{[UNIV]}\\
\bottomrule
\end{tabularx}

\caption{Language labeling schema.}
\label{tab:language-labels}
\end{table}
\begin{table*}[t]
\centering
\small
\setlength{\tabcolsep}{3pt}
\renewcommand{\arraystretch}{1.05}

\begin{tabular}{
  @{}
  >{\raggedright\arraybackslash}p{1.7cm}
  l
  l
  r
  @{}
}
\toprule
\textbf{Category} &
\textbf{Function} &
\textbf{Example} &
\textbf{\%} \\
\midrule

Metalinguistic
& \texttt{F1}. Lexical gap signal
& I can't remember the English word for \textit{종 [species]}. & 3.1\\

& \texttt{F2}. L2 equivalent request
& What do you say \textit{사춘기 [puberty]} in English? & 4.0\\

& \texttt{F3}. L1 and L2 comparison
& Is \textit{역할 놀이 [role playing]} ``role playing'' in English? & 0.8\\

& \texttt{F4}. Self-directed talk
& taste was... \textit{뭐라 해야 해 [how should I put it]}... not so strong. & 1.5\\

\midrule

\multirow[t]{3}{1.8cm}{\raggedright Meaning management}
& \texttt{F5}. Meaning restatement in L1/L2
& the \textit{목적 [goal]}, goal of the club & 2.6\\

& \texttt{F6}. L2 meaning clarification in L1
& some bad smell, \textit{비린내 [fishy smell]} & 7.3\\

& \texttt{F7}. L1 expression reformulation in L2
& \textit{한 바퀴 [one full turn]}, one cycle, one circle & 9.9\\

\midrule

Interactional
& \texttt{F8}. Interlocutor correction
& \textit{게장 [marinated crab]} is crab. Not dog. & 2.8\\

& \texttt{F9}. Interlocutor knowledge check
& Do you know about \textit{설날 [Korean Lunar New Year]}? & 4.2\\

\midrule

Content
& \texttt{F10}. Affective expression
& I feel so \textit{섭섭하다 [hurt/disappointed]}. & 4.1\\

& \texttt{F11}. Cultural reference
& eat \textit{떡국 [rice cake soup]} and do \textit{제사 [ancestral rites]}. & 44.4\\

\midrule

Unmarked
& \texttt{F12}. L1 insertion without explicit signal
& I'm trying to write a \textit{자기소개서 [personal statement]}. & 27.8\\

\midrule

Others
& \texttt{F13}. Others (\textit{e.g.}, quotation)
& a part that says ``\textit{우리 같이 저물자 [Let's fade away together]}'' & 0.3\\

\bottomrule
\end{tabular}

\caption{Taxonomy of learner CSW functions in \dataset. Percentages reflect instances labeled with each function.}
\label{tab:csw_taxonomy}
\end{table*}

\paragraph{Token language labeling.}
We split each utterance on whitespace and assigned each token one of six labels: (1) English (\texttt{EN}), general English words; (2) English borrowings (\texttt{EN-b}), Korean-origin words whose romanized forms are established in the Cambridge or Merriam-Webster Dictionary (\textit{e.g.}, kimchi); (3) Korean (\texttt{KO}), general Korean words; (4) Named entities (\texttt{NE-*}), such as people's names, locations and organizations~\cite{derczynski-etal-2017-results}; (5) Korean--English mixed (\texttt{MIX}); or (6) anonymized (\texttt{ANM}) (Table~\ref{tab:language-labels}).

An utterance is \textit{code-switched} if it includes both English and Korean, i.e., if it contains at least one \texttt{KO} or \texttt{MIX} token among \texttt{EN} tokens. Note that we exclude cases where \texttt{EN-b} or \texttt{NE-*} tokens appear alongside \texttt{EN} tokens, following \citet{poplack1988social} and \citet{molina-etal-2016-overview}.

\paragraph{Translation.}
We provide English translations of code-switched Korean instances as supplementary metadata, generated with \texttt{GPT-6 Luna}~\cite{openai2026gpt6luna} and manually verified (Prompt in App~\ref{app:post_translation}).

\subsection{CSW Function Annotation}
\label{sec:data_csw_function}

Three authors examined code-switched utterances to identify recurring functions and iteratively refined the taxonomy through discussion.
The final taxonomy comprises 13 functions across six categories (Table~\ref{tab:csw_taxonomy}).
We used multi-label annotation because a single CSW instance can serve multiple functions~\cite{belani-flanigan-2023-automatic}.
The final taxonomy comprised six categories, described in detail in App.~\ref{app:csw_taxonomy_detail}:
\begin{itemize}[topsep=0pt, itemsep=0pt]
\item \textit{Metalinguistic}: Switches that shift attention away from the ongoing conversational content to the learner's language production.
\item \textit{Meaning management}: Switches that draw on both languages to convey or clarify the learner's intended meaning.
\item \textit{Interactional}: Switches that address the interlocutor's knowledge or interpretation of what the learner has said.
\item \textit{Content}: Switches used to convey emotional meanings or culturally specific concepts as part of the conversational content.
\item \textit{Unmarked}: L1 insertions without an explicit signal indicating the purpose of the switch.
\item \textit{Other}: Infrequent functions not captured by the existing function labels.
\end{itemize}

One author initially annotated all CSW instances.
A second annotator independently annotated a random 10\% subset.
Treating each distinct, unordered label set as a single nominal category, we obtained substantial agreement with Cohen's $\kappa = 0.769$.
The second annotator then reviewed the existing annotations for the remaining instances.
The two annotators discussed and resolved disagreements across the full dataset to finalize the annotations.

\begin{table}[htbp]
\centering
\small
\setlength{\tabcolsep}{3pt}
\begin{tabular}{lrrr}
\toprule
& \textbf{Total} & \textbf{KA.} & \textbf{EO.} \\
\midrule
Learners             & 39     & 20    & 19    \\
Dialogues            & 675    & 336   & 339   \\
Total utterances   & 19,655  & 8,809 & 10,846 \\
Learner utterances   & 9,509  & 4,244 & 5,265 \\
Learner audio (h)    & 43.9   & 21.6  & 22.3 \\
\midrule
CSW dialogues      & 247    & 171   & 76   \\
CSW dialogues (\%)  & 36.6   & 50.9 & 22.4  \\
CSW utterances       & 686    & 508   & 178   \\
CSW utterances (\%)  & 7.2   & 12.0 & 3.4  \\
CSW Learner audio (h) & 4.9   & 3.9 & 1.0  \\
\bottomrule
\end{tabular}
\caption{\dataset statistics.}
\label{tab:dataset_stats}
\end{table}

\subsection{Final Dataset}
\label{sec:data_final}
Our final dataset comprises two components:
(1) a learner--LLM dialogue corpus with 675 dialogues, 19.6K utterances, and 43.9 hours of learner speech, including 247 (36.6\%) code-switched dialogues.
(2) an ASR evaluation set containing 4.9 hours of English--Korean code-switched learner speech with manually verified transcripts.
See Table~\ref{tab:dataset_stats} for details.

\section{Language Learners' CSW}
\subsection{Structural and Morphosyntactic Patterns}
We classified switching types using token-level language labels and analyzed Korean instances with \texttt{Kiwi}~\cite{lee2024kiwi}, a Korean morphological analyzer that provides morpheme-level POS tags.

\paragraph{Short, noun centered L1 insertions.}
Among code-switched instances, 96.2\% were intrasentential, 2.5\% intersentential, and 1.3\% intraword.
Intra-sentential Korean instances averaged 1.42 words.
Kiwi identified common nouns in 86.4\% of these spans, verb stems in 8.5\%, and adjective/descriptive-verb stems in 5.0\%.
These results show that learner CSW predominantly involved short, noun-centered Korean insertions within English sentences.

\paragraph{Cross-linguistic morphosyntactic integration.}
Learners integrated Korean and English elements at both syntactic and morphological levels.
Korean predicates in \textit{-다} form often appeared in English constructions, as in \textit{I want to 절약하다 [save] some money}, while Korean nouns occurred with English articles, as in \textit{a 자기소개서 [personal statement]}.
Korean adnominal modifiers preceded English nouns, as in \textit{입체적인 [multidimensional] character}.
Within words, Korean nouns took English possessive \textit{'s} and plural \textit{-s} markers, as in \textit{김치찌개 [kimchi stew]'s spicy taste} and \textit{송편 [songpyeon]s}.
These examples show how Korean elements were incorporated into English syntactic structures and combined with English morphology.

\subsection{Function Patterns}
Learners used CSW for diverse functions, as summarized in Table~\ref{tab:csw_taxonomy}.
Because annotations were multi-label, function percentages are not mutually exclusive, and sum to over 100\%.

\paragraph{Cultural references.}
Cultural reference (\texttt{F11}) was the most frequent function.
Conversations about daily life frequently involved topics such as food and holidays, prompting references to Korean foods (\textit{e.g.}, \textit{떡볶이 [tteokbokki]}) and cultural traditions (\textit{e.g.}, \textit{설날 [Korean Lunar New Year]} and \textit{제사 [ancestral rites]}).
This function frequently co-occurred with L2 reformulation (\texttt{F7}) as learners explained these concepts in English, and interlocutor knowledge checks (\texttt{F9}), as they asked whether the AI partner was familiar with them.
These patterns suggest that cultural references involved not only naming Korean concepts, but also elaborating on their meanings and negotiating a shared understanding with the interlocutor.

\paragraph{Frequent unmarked L1 insertions.}
Learners frequently inserted Korean expressions without explicitly signaling their purpose (\texttt{F12}, 27.8\%), whereas explicit lexical gap signals (\texttt{F1}, 3.1\%) and requests for L2 equivalents (\texttt{F2}, 4.0\%) were less common.
In some instances, we observed hesitation, fillers, rising intonation, or embarrassed laughter around the switch.
Although these cues may suggest difficulty in retrieving an L2 expression, they do not establish that a lexical gap motivated the switch.
Unmarked insertions should therefore not be interpreted uniformly as evidence of lexical difficulty.

\paragraph{Constructing meaning across languages.}
Learners used CSW not only to substitute for unavailable L2 vocabulary but also to use both languages to progressively specify their intended meanings.
Some followed an approximate English expression with Korean clarification (\texttt{F6}), while others introduced a Korean expression before reformulating it in English through approximation, circumlocution, or word coinage (\texttt{F7}).
These sequences reveal both what learners intend to convey and how they attempt to express it in L2, offering an AI partner clues about what learners can already express and where further support may be useful.
Supporting these attempts would allow learners to draw on their full linguistic repertoire while developing their L2 expression.

\paragraph{CSW with an English speaking partner.}
Even when the partner presented itself as communicating only in English, learners used CSW in 22.4\% of their conversations.
In the EO setting, CSW mainly served cultural references (\texttt{F11}, 78.3\%), L2 reformulation (22.2\%), and L1 clarification (7.2\%). 
These patterns suggest that learners naturally draw on their L1 to convey culturally specific content and manage meaning, even with an English-only partner. 
Language-learning systems should therefore be intentionally designed with learners' L1 resources and their communicative functions in mind, even for English-only practice.

\subsection{Learner Responses to AI Partner Support Following CSW}
We examined how learners responded to AI partner responses following CSW. Examples 1–4 illustrate four observed patterns: \textit{exact reuse}, \textit{adapted reuse}, \textit{correction (of an AI misunderstanding)}, and \textit{continued use of Korean}.

\begin{tcolorbox}[breakable,
    colback=gray!5!white,   
    colframe=gray!50!gray, 
    fonttitle=\bfseries,   
    arc=1mm,               
    boxrule=0.5pt,         
    fontupper=\small,
        boxsep=1pt,           
    left=2pt,             
    right=2pt,            
    top=2pt,              
    bottom=2pt,            
]
\textbf{Example 1: Exact Reuse}\\
\texttt{[L]:} Yes, I'm trying to write a \textbf{자기소개서} to apply company and it is quite hard.\\
\texttt{[AI]:} Oh, you're working on a \textbf{자기소개서}? \textbf{In English, it's called a "self-introduction letter"} or "personal statement." It can be challenging to express yourself well. What part are you finding most difficult?\\
\texttt{[L]:} Uh.. I feel my... \textbf{self-introduction letter} is too vague.
\end{tcolorbox}

\begin{tcolorbox}[breakable,
    colback=gray!5!white,   
    colframe=gray!50!gray, 
    fonttitle=\bfseries,   
    arc=1mm,               
    boxrule=0.5pt,         
    fontupper=\small,
        boxsep=1pt,           
    left=2pt,             
    right=2pt,            
    top=2pt,              
    bottom=2pt,            
]
\textbf{Example 2: Adapted Reuse}\\
\texttt{[L]:} What is... What is \textbf{굽다} in English?\\
\texttt{[AI]:} For \textbf{굽다, you can say 'grill'} or 'bake' in English, depending on the context. Do you like to grill or bake food?\\
\texttt{[L]:} I like \textbf{grilled} meat and \textbf{grilled} pork ...
\end{tcolorbox}

\begin{tcolorbox}[breakable,
    colback=gray!5!white,   
    colframe=gray!50!gray, 
    fonttitle=\bfseries,   
    arc=1mm,               
    boxrule=0.5pt,         
    fontupper=\small,
        boxsep=1pt,           
    left=2pt,             
    right=2pt,            
    top=2pt,              
    bottom=2pt,            
]
\textbf{Example 3: Correction}\\
\texttt{[L]:} We usually eat cold noodle with vinegar and \textbf{겨자}.\\
\texttt{[AI]:} That sounds delicious! Cold noodles with vinegar must be refreshing. Did you mean \textbf{교자}? They're tasty dumplings...\\
\texttt{[L]:} No, \textbf{I didn't mean 교자. I said 겨자, which is like spicy yellowish thing}, if you know what I mean.
\end{tcolorbox}

\begin{tcolorbox}[breakable,
    colback=gray!5!white,   
    colframe=gray!50!gray, 
    fonttitle=\bfseries,   
    arc=1mm,               
    boxrule=0.5pt,         
    fontupper=\small,
        boxsep=1pt,           
    left=2pt,             
    right=2pt,            
    top=2pt,              
    bottom=2pt,            
]
\textbf{Example 4: Continued Use of Korean}\\
\texttt{[L]:} ...next week is \textbf{설날}. Do you know about \textbf{설날}?\\
\texttt{[AI]:} Yes, I know a little about \textbf{설날}! \textbf{It's the Korean New Year, right?} It's a time for family gatherings and traditional food. What do you usually do to celebrate?\\
\texttt{[L]:} ...However, in the \textbf{설날}, I usually eat traditional Korean food like 떡국 or 전.
\end{tcolorbox}

Examples 1 and 2 show immediate reuse of an AI-provided expression, either verbatim or in an adapted form. 
Example 3 illustrates meaning negotiation, as the learner corrects the AI partner's misunderstanding and clarifies the intended meaning.
Such misunderstandings may arise from ASR errors or the LLM's limited L1 knowledge.
In Example 4, the learner continues using the Korean term despite an English reformulation. This may reflect a preference for its cultural specificity or nuance, or simply a choice not to reuse the English expression.
Together, these examples show that learners respond to AI partner support in various ways, beyond immediate reuse of an English equivalent.

Learners also occasionally referred to lexical assistance from earlier interactions. One learner recalled prior assistance while struggling to retrieve an expression: ``I love \textit{방어 [yellowtail]}, but I forgot how to express in Korean, especially you teach me a last day, but, and...''. Another asked for a previously supplied equivalent: ``So what did you say \textit{본질적인 [fundamental]} is in English?''.
These instances suggest that learners' engagement with lexical support can extend beyond the immediate exchange, recall receiving assistance without retrieving the expression itself.
However, because these patterns are observational, they should not be interpreted as evidence of vocabulary acquisition or retention.

\subsection{CSW Learner Simulation}

We examine whether LLM-simulated learners produce similar CSW patterns as in \dataset.
We use \texttt{GPT-6.1 Sol}~\cite{openai2026gpt61} and \texttt{Gemini-3.8-Flash}~\cite{deepmind2026gemini38flash}, each playing both the learner and the AI partner. We compare persona prompting (Base) with added taxonomy and examples (+Tax), generating 30 dialogues per model--setting--condition combination (240 total), each capped at 15 turns to approximate the average dialogue length in \dataset (prompts in App.~\ref{app:simulation}). We compare CSW frequency per learner utterance and structure and functions per CSW instance, allowing multiple function labels. Simulated functions are annotated using \texttt{GPT-6.1-Sol}, which achieved 72.3\% classification accuracy on real-learner data (App.~\ref{app:csw_function_classification}). 

\begin{table}[t]
    \centering
    \footnotesize
    \setlength{\tabcolsep}{2pt}
    \renewcommand{\arraystretch}{1.1}

    \begin{tabularx}{\linewidth}{
        @{}>{\raggedright\arraybackslash}Xrrrrr@{}
    }
        \toprule
        & & \multicolumn{2}{c}{Base}
            & \multicolumn{2}{c}{+Taxonomy} \\
        \cmidrule(lr){3-4}
        \cmidrule(lr){5-6}
        & \dataset & GPT & Gemini & GPT & Gemini \\
        \midrule
        CSW utterances
        & 7.2
        & \cswpos{69}{76.4} & \cswpos{66}{73.3}
        & \cswpos{62}{69.6} & \cswpos{68}{75.1} \\
        \midrule
        Intersentential
        & 2.5
        & \cswpos{48}{50.7} & \cswpos{29}{31.1}
        & \cswpos{42}{44.1} & \cswpos{8}{10.6} \\
        Intrasentential
        & 96.2
        & \cswneg{47}{49.0} & \cswneg{28}{68.1}
        & \cswneg{40}{55.8} & \cswneg{7}{89.2} \\
        Intraword
        & 1.3
        & \cswneg{1}{0.3} & \cswneg{1}{0.8}
        & \cswneg{1}{0.2} & \cswneg{1}{0.3} \\
        \midrule
        F1. Lexical Gap
        & 3.1
        & \cswneg{3}{0.3} & \cswneg{3}{0.1}
        & \cswneg{2}{1.3} & \cswneg{2}{1.5} \\
        F2. Request
        & 4.0
        & \cswneg{3}{1.1} & \cswneg{4}{0.1}
        & \cswpos{2}{5.8} & \cswpos{3}{7.4} \\
        F3. Comparison
        & 0.8
        & \cswpos{0}{1.1} & \cswneg{0}{0.6}
        & \cswpos{6}{6.3} & \cswpos{2}{2.5} \\
        F4. Self-talk
        & 1.5
        & \cswpos{1}{2.6} & \cswneg{1}{0.6}
        & \cswpos{8}{9.2} & \cswpos{4}{5.4} \\
        F5. Restatement
        & 2.6
        & \cswpos{3}{5.9} & \cswpos{21}{24.0}
        & \cswpos{4}{6.9} & \cswpos{23}{26.0} \\
        F6. Clarification
        & 7.3
        & \cswneg{5}{2.0} & \cswneg{6}{1.2}
        & \cswneg{6}{1.1} & \cswneg{3}{4.3} \\
        F7. Reformulation
        & 9.9
        & \cswneg{2}{8.0} & \cswpos{8}{18.0}
        & \cswpos{8}{18.3} & \cswpos{10}{20.1} \\
        F8. Correction
        & 2.8
        & \cswneg{3}{0.0} & \cswneg{3}{0.0}
        & \cswneg{3}{0.0} & \cswneg{3}{0.0} \\
        F9. Check
        & 4.2
        & \cswneg{4}{0.0} & \cswneg{4}{0.7}
        & \cswneg{3}{0.8} & \cswneg{3}{1.7} \\
        F10. Affect
        & 4.1
        & \cswpos{13}{17.2} & \cswpos{13}{17.1}
        & \cswpos{23}{26.9} & \cswpos{12}{16.3} \\
        F11. Culture
        & 44.4
        & \cswneg{43}{1.1} & \cswneg{28}{16.4}
        & \cswneg{36}{8.6} & \cswneg{24}{20.8} \\
        F12. Insertion
        & 27.8
        & \cswpos{28}{56.0} & \cswpos{5}{32.7}
        & \cswneg{2}{26.1} & \cswneg{15}{13.2} \\
        F13. Others
        & 0.3
        & \cswpos{8}{7.8} & \cswpos{4}{3.8}
        & \cswpos{8}{8.1} & \cswpos{3}{3.1} \\
        \bottomrule
    \end{tabularx}
    \caption{Learner CSW rates in \dataset and simulated learners. Cell color shows the \colorbox{over!20}{positive} or \colorbox{under!20}{negative} percentage-point deviation from \dataset.}
    \label{tab:csw_learner_simulation_pooled}
\end{table}

Under persona-only prompting, simulated learners differ substantially from real learners in CSW frequency, structure, and functions. They code-switch in 76.4\% of learner turns in GPT and 73.3\% in Gemini, compared with 7.2\% in \dataset. Real learners predominantly use intrasentential CSW (96.2\%), whereas its share is substantially lower in GPT (49.0\%) and Gemini (68.1\%). Both models underproduce cultural references and overproduce unmarked insertions. Interlocutor correction is absent from both models' simulations, and knowledge checks are absent or rare.

Providing our taxonomy and examples narrows some of these gaps. The share of intrasentential CSW increases to 55.8\% in GPT and 89.2\% in Gemini, bringing both models closer to the real-learner pattern. Knowledge checks and cultural references also become more frequent. However, these improvements are selective. Interlocutor correction remains absent, and overall switching frequency and functional distributions still differ substantially from the real-learner patterns.

Together, these results show that persona-only prompting produces learner simulations that differ substantially from real learners' CSW patterns, highlighting the value of \dataset\ as an empirical reference for evaluating and calibrating learner simulations. Providing our taxonomy and examples brings CSW structure and selected functions closer to the real-learner patterns, demonstrating their usefulness for guiding learner simulations. However, persistent gaps in overall switching frequency and the distribution of CSW functions suggest that future work should explore more fine-grained guidance to better align simulations with real learner behavior.

\section{Learner CSW ASR Evaluation}
\subsection{Experimental Setup}

\paragraph{Datasets.}
\datasetasr comprises 4.9 hours (668 utterances) of Korean--English code-switched learner speech with human-verified transcripts.
We also evaluate the models on HiKE~\cite{paik-etal-2026-hike}, a Korean--English CSW benchmark not specifically designed for learner speech, and the Korean and English subsets of FLEURS~\cite{conneau2023fleurs}, a monolingual ASR benchmark.

\paragraph{STT Models.}
We evaluate three proprietary models and eight open models.
The proprietary models are \texttt{Gemini-3.5-Transcribe}~\cite{deepmind2026gemini35audio}, \texttt{GPT-4o Transcribe}, and \texttt{GPT-4o Mini Transcribe}~\cite{openai2024gpt4otrans}.
The open models included \texttt{Qwen3-ASR-1.7B} and \texttt{Qwen3-ASR-0.6B}~\cite{shi2026qwen3asr}, as well as six \texttt{Whisper} variants: \texttt{large-v3}, \texttt{large-v2}, \texttt{medium}, \texttt{small}, \texttt{base}, and \texttt{tiny}~\cite{radford2023robust}.

\paragraph{Evaluation Metrics.}
For Korean--English code-switched speech, we report Mixed Error Rate (MER)~\cite{paik-etal-2026-hike} and Point of Interest Error Rate (PIER)~\cite{ugan2025pier}.
For monolingual speech, we report Character Error Rate (CER) on the Korean subset of FLEURS and Word Error Rate (WER) on the English subset.
Lower values indicate better performance for all metrics.

\subsection{Results}

\begin{table}[htbp]
    \centering
    \small
    \setlength{\tabcolsep}{2.5pt}
    \begin{tabular}{@{}lrrrrrr@{}}
        \toprule
        \multirow{2}{*}{Model}
        & \multicolumn{1}{c}{EN}
        & \multicolumn{1}{c}{KO}
        & \multicolumn{2}{c}{HiKE}
        & \multicolumn{2}{c}{\datasetasr} \\
        \cmidrule(lr){2-2}
        \cmidrule(lr){3-3}
        \cmidrule(lr){4-5}
        \cmidrule(lr){6-7}
        & \multicolumn{1}{c}{WER}
        & \multicolumn{1}{c}{CER}
        & \multicolumn{1}{c}{MER}
        & \multicolumn{1}{c}{PIER}
        & \multicolumn{1}{c}{MER}
        & \multicolumn{1}{c}{PIER} \\
        \midrule
        Gemini 3.5 Tr.
        & 3.7 & 4.2 & \textbf{9.2} & \textbf{18.3} & \textbf{20.5} & 80.5 \\
        GPT-4o Tr.
        & \textbf{3.5} & \textbf{4.0} & 22.6 & 28.7 & 37.9 & 60.7 \\
        GPT-4o Mini Tr.
        & 4.0 & 5.5 & 26.0 & 32.3 & 22.9 & \textbf{57.8} \\
        \midrule
        Qwen3-ASR 1.7B
        & 4.1 & 4.5 & 24.2 & 42.8 & 22.9 & 110.5 \\
        Qwen3-ASR 0.6B
        & 5.2 & 5.8 & 23.6 & 39.1 & 26.6 & 124.2 \\
        Whisper large-v3
        & 5.0 & 4.9 & 34.7 & 45.4 & 49.8 & 81.8 \\
        Whisper large-v2
        & 4.9 & 5.0 & 34.1 & 39.6 & 74.4 & 75.7 \\
        Whisper medium
        & 5.4 & 5.4 & 31.4 & 36.9 & 97.9 & 114.9 \\
        Whisper small
        & 7.0 & 7.7 & 35.3 & 44.7 & 81.7 & 167.6 \\
        Whisper base
        & 11.2 & 12.9 & 106.4 & 109.5 & 220.2 & 158.3 \\
        Whisper tiny
        & 14.7 & 19.0 & 84.2 & 86.5 & 206.1 & 230.0 \\
        \bottomrule
    \end{tabular}
    \caption{ASR error rates (\%, $\downarrow$) on \datasetasr.}
    \label{tab:asr-results}
\end{table}

The results suggest that ASR for code-switched learner speech remains challenging even for models with low monolingual error rates (Table~\ref{tab:asr-results}).
Most models exhibit higher MER and PIER on \datasetasr than on HiKE, suggesting greater difficulty recognizing learner's CSW speech.
The gap is more pronounced for PIER and widened among smaller Whisper models.
These findings highlight ASR accuracy on code-switched learner speech as a potential bottleneck for speech-based educational systems that support learner CSW.

\section{Discussion}
\subsection{Toward AI Tutors That Understand and Support Learner Code-Switching}
Our findings suggest that supporting learner CSW requires an understanding of the function of a switch within the ongoing interaction.
The diverse functions of observed CSW indicate that a switch should not be treated uniformly as evidence of L2 deficiency or as a request for translation. 
Some switches explicitly signal lexical difficulty or request an L2 equivalent, whereas others clarify intended meaning, check the interlocutor's knowledge, express affect, or introduce culturally specific content. 
A Korean cultural reference, for example, may convey a concept for which an English equivalent is not the learner's primary communicative need. 
AI tutors should thus consider learners' communicative intent and existing linguistic resources when deciding whether and how to intervene.

Recognizing these functions also depends on how well the conversational AI systems process learner speech. 
Accurate recognition of code-switched speech provides access to learners' intended content, while features of spoken interaction, such as hesitation, repair, and prosody, may offer contextual clues about possible difficulty or uncertainty.
Although our ASR evaluation did not assess these additional cues, it shows that even transcribing code-switched learner speech remains challenging, particularly the CSW portions.

Supporting learner CSW thus requires coordination between speech recognition, contextual interpretation, and pedagogical response. 
Rather than automatically translating or correcting every L1 expression, a conversational tutor could first identify the likely function of the switch in context and then decide whether linguistic support is warranted. 
For instance, an explicit request for an L2 equivalent may call for direct lexical assistance, whereas a cultural reference may instead call for acknowledgement, clarification, or continuation of the conversation without correction.
This function-sensitive approach would accommodate learners' multilingual resources as part of the interaction while providing support suited to their communicative needs.

\subsection{From Code-Switching to Opportunities for Linguistic Support}
The observed reuse and adaptation of AI-provided expressions suggest how CSW can create opportunities for contextualized linguistic support.
For conversational tutoring systems, these patterns highlight the potential of providing lexical assistance when it becomes relevant to what learners are trying to express.
References to earlier assistance also suggest that learners' engagement with this support can extend beyond the immediate exchange.
However, neither immediate reuse nor recall of prior assistance establishes vocabulary acquisition or retention.
These findings motivate developing scaffolding strategies that extend beyond immediate uptake.
Future work could examine how such support can be tailored to the functions of learner CSW, and whether it promotes independent L2 use and retention over time.

\section{Conclusion}
We introduce \dataset, a corpus of 675 spoken learner--AI dialogues, a taxonomy of learner CSW functions, and a 4.9-hour ASR evaluation set with manually verified transcripts.
Our findings show that learner CSW serves diverse communicative needs and creates opportunities for lexical uptake, while remaining challenging for current ASR models.
These findings highlight the need for conversational language-learning systems that can recognize learner CSW, interpret its function in context, and respond appropriately.

\section{Limitations}
Our dataset captures interactions involving 39 Korean undergraduate learners of English. Therefore, the observed CSW patterns may not generalize to other language pairs or learner populations.
Also, our functional annotations characterize observable communicative functions based on utterances and interactional context, but do not establish learners' underlying motivations for switching.
The transcription-cleaning pipeline may miss code-switches transcribed as English by all ASR systems, and residual transcription errors may remain in unflagged utterances that were not manually verified. Such omissions and errors may affect estimates of CSW frequency and expression reuse.
Finally, our analysis of expression reuse describes learners' responses to AI-provided support rather than demonstrating vocabulary acquisition or retention.

\section{Ethical Considerations}
The study was approved by our institutional review board (IRB).
Participants provided written consent to participate in the study. Participation was voluntary, and participants could withdraw at any time.
Each participant received KRW 100,000 (approximately USD 72) for approximately three hours of participation over six weeks, although the actual duration varied across participants.
To reduce identification risks, we remove personally identifiable information (PII) from transcripts and exclude audio recordings containing PII from public dataset. However, released audio may still permit speaker identification through voice characteristics.
We release the dataset for research purposes only to support further research on LLMs in education.

\bibliography{custom}

\clearpage
\section*{Appendix}
\appendix
\label{sec:appendix}
\section{Participant Demographics}
\label{app:participant_info}
The study includes 39 participants aged 18–30 years (M = 22.74, SD = 2.96), comprising 18 women and 21 men. Each participant reported their TOEFL iBT Speaking or IELTS scores. Self-reported IELTS Speaking scores are mapped to the TOEFL iBT Speaking 0–30 scale using the conversion procedure and categorized according to ETS’s published proficiency thresholds.\footnote{\url{https://www.ets.org/toefl/teachers-advisors-agents/ibt/scores.html}} The resulting scores have a mean of 18.08 (SD = 4.93; range = 6–27).

\begin{table}[h]
\centering
\small
\setlength{\tabcolsep}{3pt}
\begin{tabular}{@{}lrrr@{}}
\toprule
 & Overall & KA. & EO. \\
\midrule
Participants, $n$ & 39 & 20 & 19 \\
Age, $M$ ($SD$) & 22.74 (2.96) & 23.40 (3.10) & 22.05 (2.72) \\
Age range & 18--30 & 18--30 & 18--26 \\
Women / men, $n$ & 18 / 21 & 12 / 8 & 6 / 13 \\
\midrule
\multicolumn{4}{@{}l}{\textit{Speaking proficiency level}} \\
Below Basic & 3 & 2 & 1 \\
Basic & 9 & 4 & 5 \\
Low-Intermediate & 9 & 5 & 4 \\
High-Intermediate & 14 & 7 & 7 \\
Advanced & 4 & 2 & 2 \\
\bottomrule
\end{tabular}
\caption{Participant characteristics and speaking proficiency
levels overall and by group.}
\label{tab:participant-characteristics}
\end{table}

\section{LLM-based Speaking Practice System}

\subsection{ASR Implementation Details}
\label{app:system_csw_asr}
We use \texttt{Whisper large-v2}~\cite{radford2023robust} with speech-based in-context learning~\cite{wang2024whisper} to transcribe learner utterances.

To construct the in-context examples, two speakers (one male and one female) recorded 20 short, clearly articulated CSW samples: five Korean-matrix samples with English insertions, five English-matrix samples with Korean insertions, and ten samples reflecting EFL learners' code-switching. The samples cover code-switching functions identified in a classroom observation study~\cite{yoon2023code}. We iteratively refine the samples until the base \texttt{Whisper large-v2} model consistently transcribed them accurately.

At runtime, we retrieve the in-context sample most similar to the current learner utterance using vector embedding~\cite{baevski2020advances} similarity and use the selected sample for speech-based in-context learning.

\begin{tcolorbox}[
breakable,
top=1pt,
left=1pt,
right=1pt,
bottom=1pt,
colback=white,
fontupper=\ttfamily,
fonttitle=\bfseries,
title={CSW In-Context Learning Samples}
]
\small
$<$Korean-Matrix Samples$>$\\
날씨가 너무 unpredictable 한 것 같아.\\
그 영화 진짜 funny 했어.\\
지난 강의가 정말 meaningful 했어.\\
그 사람 상당히 persuasive 하더라.\\
어제 모임 참석자들이 모두 friendly 해서 좋았어.\\

$<$English-Matrix Samples$>$\\
I think the weather is 쌀쌀해 today.\\
That movie was really 웃겼어.\\
The last lecture was really 유익했어.\\
That person was 상당히 persuasive.\\
Everyone at yesterday's meeting was 친절했어.\\

$<$EFL Learners' CSW Samples$>$\\
My camera is 망가졌어.\\
I think that is somewhat... 모순적이다.\\
I dropped my phone, so 망가졌다가 영어로 기억이 안 나네.\\
I don't know how to say 설레다 in English.\\
I enjoyed watching the farmers sing 모내기 소리.\\
떡볶이 is a popular street food.\\
I like Lord of the Rings. 반지의 제왕.\\
You can do it! 화이팅해!\\
그러네, that's true.\\
뭐였지? I forgot.
\end{tcolorbox}

\subsection{LLM-based Response Construction}
\label{app:response_construction}

To guide the design of the CSW tutor, we conducted an exploratory formative interview with eight South Korean students. Participants read five dialogue scenarios involving Korean CSW during English speaking practice and wrote the responses they would like to receive from an AI partner, using English, Korean, or both. The scenarios covered five functions of code-switching drawn from prior work, as shown below. [L] denotes the learner and [AI] the AI partner.

\begin{tcolorbox}[breakable, colback=gray!5!white, colframe=gray!50!gray, arc=1mm, boxrule=0.5pt, fontupper=\small, boxsep=1pt, left=2pt, right=2pt, top=2pt, bottom=2pt]
\textbf{Scenario 1: Compensating for limited L2 knowledge~\cite{kim2012functions}}\\
\texttt{[L]:} I broke my phone yesterday.\\
\texttt{[AI]:} Oh that's so bad. How did that happen?\\
\texttt{[L]:} I dropped my phone in the pool, and now the camera is, um.. like \textbf{뿌옇다}?\\
\texttt{[AI]:}
\end{tcolorbox}

\begin{tcolorbox}[breakable, colback=gray!5!white, colframe=gray!50!gray, arc=1mm, boxrule=0.5pt, fontupper=\small, boxsep=1pt, left=2pt, right=2pt, top=2pt, bottom=2pt]
\textbf{Scenario 2: Metalinguistic use~\cite{yoon2023code}}\\
\texttt{[L]:} I broke my phone yesterday.\\
\texttt{[AI]:} Oh that's so bad. How did that happen?\\
\texttt{[L]:} I dropped my phone in the pool, and now the camera is, um.. \textbf{뿌옇다가 영어로 뭐였지?}\\
\texttt{[AI]:}
\end{tcolorbox}

\begin{tcolorbox}[breakable, colback=gray!5!white, colframe=gray!50!gray, arc=1mm, boxrule=0.5pt, fontupper=\small, boxsep=1pt, left=2pt, right=2pt, top=2pt, bottom=2pt]
\textbf{Scenario 3: Conveying culture-specific content~\cite{kim2012functions}}\\
\texttt{[AI]:} What would you recomend doing in Korea?\\
\texttt{[L]:} Going to a \textbf{찜질방} would be great! You should also drink \textbf{식혜} there.\\
\texttt{[AI]:}
\end{tcolorbox}

\begin{tcolorbox}[breakable, colback=gray!5!white, colframe=gray!50!gray, arc=1mm, boxrule=0.5pt, fontupper=\small, boxsep=1pt, left=2pt, right=2pt, top=2pt, bottom=2pt]
\textbf{Scenario 4: Clarifying one's message~\cite{kim2012functions}}\\
\texttt{[AI]:} What did you do last weekend?\\
\texttt{[L]:} I watched a movie, \textit{The Roundup}, \textbf{범죄도시 4}.\\
\texttt{[AI]:}
\end{tcolorbox}

\begin{tcolorbox}[breakable, colback=gray!5!white, colframe=gray!50!gray, arc=1mm, boxrule=0.5pt, fontupper=\small, boxsep=1pt, left=2pt, right=2pt, top=2pt, bottom=2pt]
\textbf{Scenario 5: Unintentional remarks~\cite{noh2015study}}\\
\texttt{[AI]:} What's your favorite thing to do in your free time?\\
\texttt{[L]:} I like to watch movies. I watched \textit{Inside Out 2} recently.\\
\texttt{[AI]:} I loved that movie too. It made me so emotional.\\
\texttt{[L]:} \textbf{그니까!} Oh, I mean, you're right, me too.\\
\texttt{[AI]:}
\end{tcolorbox}

Through thematic analysis of the responses, we identified four response strategies: indirect expression support (\textit{e.g.}, ``Oh, did your phone camera lens get foggy?''), direct expression support (\textit{e.g.}, ``You can use `foggy' for `뿌옇다' ''), requests for explanation (\textit{e.g.}, ``What’s 식혜? Can you explain that to me?''), and continuing the conversation without acknowledgment.
We used these strategies to inform the response guidelines in the Korean-aware partner prompt.

\subsection{LLM-based Response Generation Prompt}
\label{app:system_response_gen}

\begin{tcolorbox}[
breakable,
top=1pt,
left=1pt,
right=1pt,
bottom=1pt,
colback=white,
fontupper=\ttfamily,
fonttitle=\bfseries,
title={Monolingual Partner Prompt}
]
\small
You are Emily, a friendly and patient English conversation partner. Your role is to engage in casual, natural conversations on various topics such as hobbies, travel, food, and culture.\\

You can only understand English, and do not speak in Korean.
Your primary goal is to encourage English usage while making the conversation comfortable and enjoyable.
Note that this is a speaking conversation. Answer with short sentences.\\

Maintain an informal, encouraging tone to help me feel confident about using English.  
\end{tcolorbox}

\begin{tcolorbox}[
breakable,
top=1pt,
left=1pt,
right=1pt,
bottom=1pt,
colback=white,
fontupper=\ttfamily,
fonttitle=\bfseries,
title={Bilingual Partner Prompt}
]
\small
You are Emily, a friendly and patient English conversation partner. Your role is to engage in casual, natural conversations on various topics such as hobbies, travel, food, and culture.\\

You can understand both English and Korean, but do not speak in Korean. You are also familiar with Korean culture.
Your primary goal is to encourage English usage while making the conversation comfortable and enjoyable.
Note that this is a speaking conversation. Answer with short sentences.\\

When I use Korean during our conversation, follow these steps:\\  
1. Identify the Type of Korean Usage:\\
   Classify my use of Korean into one of the types below.\\  
2. Choose the Appropriate Reaction:\\ 
   Respond according to the suggested reactions for each type, while keeping the conversation natural and encouraging.\\  
3. Optional:\\
   If a Korean word does not affect the overall understanding of the conversation, you may choose not to address it explicitly.\\

Types of Korean Usage and Reactions\\

Type 1: Replacing an Unknown or Forgotten English Expression with Korean\\
Reaction 1: Indirectly introduce the correct English expression while continuing the conversation.\\  
Example:\\  
User: "I like 축구."\\  
Emily: "Oh, do you mean you like to play soccer?"\\  

Reaction 2: Directly provide the English expression.\\  
Example:\\  
User: "I like 축구."\\  
Emily: "축구 is 'soccer' in English."\\

Type 2: Directly Asking About an Unknown or Forgotten English Expression\\
Reaction: Directly provide the English expression.\\  
  Example:\\  
  User: "I like 축구. What’s 축구 in English?"\\  
  Emily: "For 축구, you can say 'soccer' in English."\\

Type 3: Describing Something Related to Korean Culture\\
Reaction 1: Directly provide the English expression.\\  
Example:\\  
User: "I learned new 동요 in school."\\  
Emily: "You can say 'nursery rhyme' or 'children's song' for 동요."\\  

Reaction 2: Ask follow-up questions to encourage users to provide a more detailed explanation of the Korean cultural term (e.g., What do you do there? What do you eat there? When do you eat that food?).\\
Example:\\  
User: "I went to 찜질방 yesterday."\\  
Emily: "What do you usually do in Jjimjilbang?"\\

After addressing the Korean usage, continue the conversation smoothly and naturally.  
Maintain an informal, encouraging tone to help me feel confident about using English. 
\end{tcolorbox}

\subsection{Speaking Practice Platform}
\label{app:system_images}


\begingroup
\captionsetup{type=figure}
\centering

\subcaptionbox{LLM speaking partner is speaking. The audio playback bar is displayed to indicate progress.\label{fig:system1}}[\columnwidth]{%
    \includegraphics[width=0.8\columnwidth]{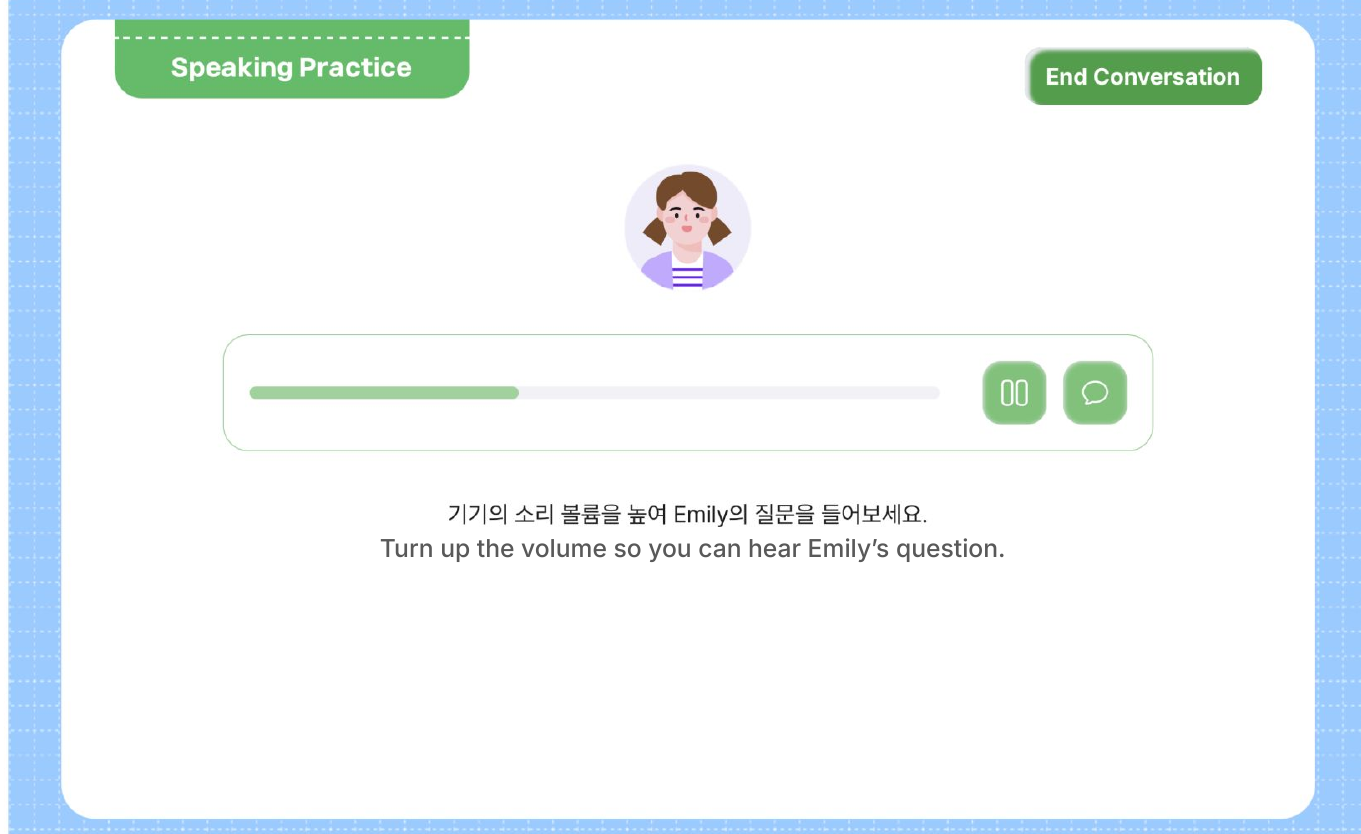}%
}
\par\medskip

\subcaptionbox{Learner is speaking. A real-time audio level meter visualizes the learner’s speech intensity. When the learner finishes speaking, pressing the Stop button ends the turn and hands the turn back to Emily.\label{fig:system2}}[\columnwidth]{%
    \includegraphics[width=0.8\columnwidth]{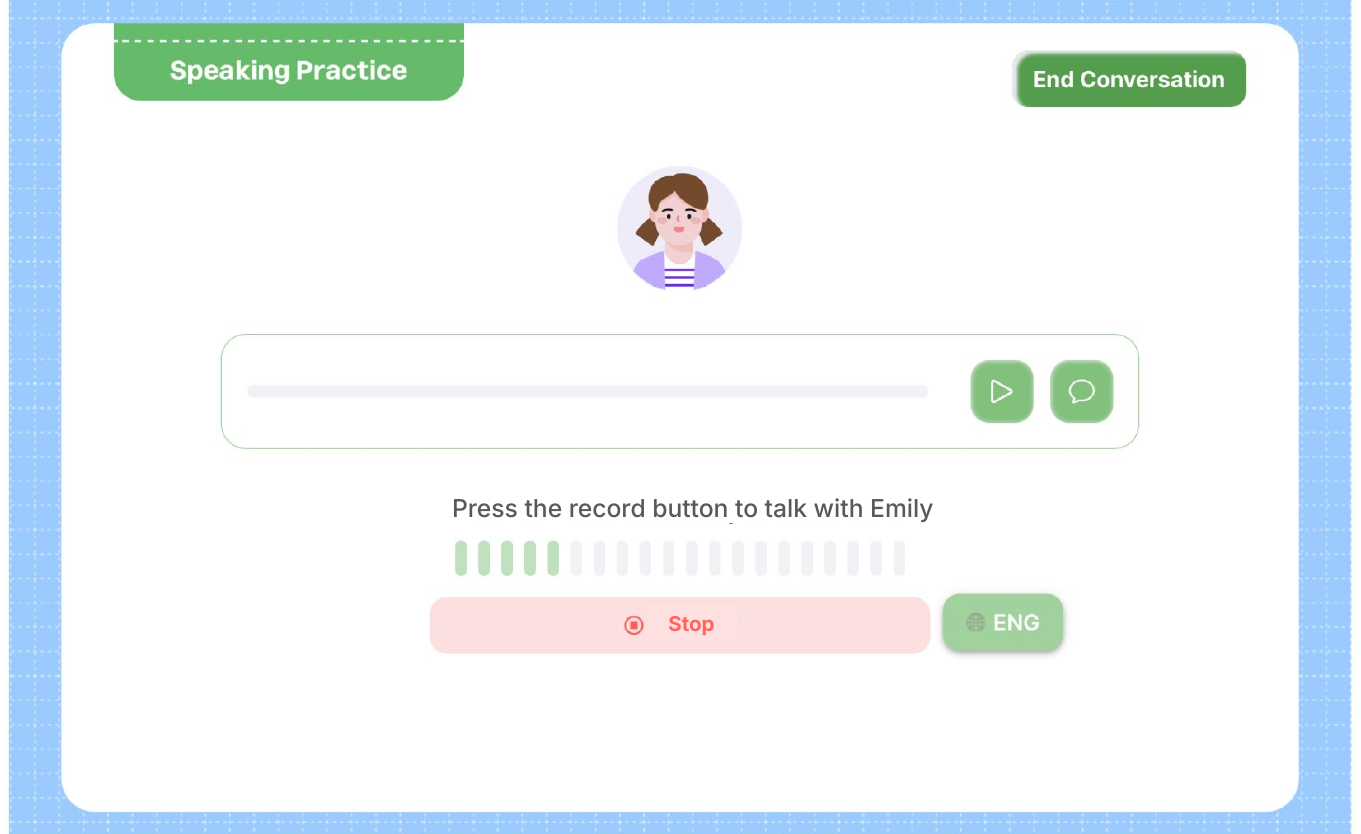}%
}


\caption{Speaking practice platform interface.}
\label{fig:system_interface}
\par
\endgroup

\section{Data Post Processing}
\subsection{CSW Detection Prompt}
\label{app:post_transcription_cleaning}

\begin{tcolorbox}[
breakable,
top=1pt,
left=1pt,
right=1pt,
bottom=1pt,
colback=white,
fontupper=\ttfamily,
fonttitle=\bfseries,
title={CSW Detection Prompt}
]
\small
You are checking ASR transcript (text).\\

Decide whether ANY of the fields below contain Hangul (Korean script) OR romanized Korean (including Korean names, titles, loanwords written in Latin letters, e.g. annyeong, gaja, Nagwa, Oncheon, Seoul, Gyeongbokgung, Sooyoung, kimchi, tteokbokki).\\

If none of the fields contain Hangul or romanized Korean, answer with "None". If there is any Hangul or romanized Korean, respond with that word/phrase/sentence. Do not include any explanation.\\

Fields:
\{fields\}
\end{tcolorbox}

\subsection{PII Anonymization Prompt}
\label{app:post_pii_anonymization}

\begin{tcolorbox}[
breakable,
top=1pt,
left=1pt,
right=1pt,
bottom=1pt,
colback=white,
fontupper=\ttfamily,
fonttitle=\bfseries,
title={PII Anonymization Prompt}
]
\small
This is a {role}'s utterance in a conversation between a student and AI conversation partner named Emily.\\ 

Please edit the provided text by removing any personally identifiable information (PII). This includes names of students, or professors, university name, company name, research lab name, places of origin, current living location, address, date, age, or anything else that may lead to identifying the student.\\

Generic indicators such as ‘the school’, ‘the hospital’, 'the laboratory', 'my hometown', should not be removed. Names of artists, philosophers, politicians, or locations that do not correspond to the student's place of origin or current location should not be removed.\\ The name 'Emily' should not be removed.\\

Examples:\\
- Name (Hi, I’m XX, you can call me XX) -> replace with [NAME]\\
- University name (I go to XX, I attend XX, I am a student in XX) -> replace with [UNIV]\\
- Company name (I am working in XX, I am interning in XX) -> replace with [COMP]\\
- Location (my hometown is XX, I live in XX, my family live in XX) -> replace with [LOC]\\
- Location (A cafe XX near my school, a restaurant XX in my neighborhood) -> replace the cafe, restaurant name with [LOC]\\
- Date (my birthday is XX) -> replace with [DATE]\\
- Age (I am XX years old) -> replace with [AGE]\\
- Other PIIs -> replace with [OTHER]\\

Ensure that the rest of the text remains unchanged, word for word.\\ Maintain the original punctuation, quotation marks, spaces and line breaks. If the text does not contain any PII, return “NONE”.
\end{tcolorbox}

\subsection{Translation Prompt}
\label{app:post_translation}

\begin{tcolorbox}[
    breakable,
    top=1pt,
    left=1pt,
    right=1pt,
    bottom=1pt,
    colback=white,
    fontupper=\ttfamily,
    fonttitle=\bfseries,
    title={Translation Prompt}
]
\small
Given a Korean phrase and a user utterance that provides context, provide a translated English phrase: One or more natural English translations or equivalents of the Korean phrase that best fit the given context.\\
If the Korean term is commonly used in English in its romanized Korean form, you may include that as an alternative.\\
Definition: A concise, one-sentence English definition explaining what the Korean phrase means in the context of the user utterance.\\
Return only valid JSON. Do not include explanations, comments, Markdown, or any text outside the JSON object.

\par\medskip
\noindent
Output format:\\
\{\\
\hspace*{1em}"translated\_english\_phrase": ["<translation 1>", "<translation 2>"],\\
\hspace*{1em}"definition": "<one-sentence contextual definition>"\\
\}

\par\medskip
\noindent
Example input:\\
Korean phrase: 설날\\
User utterance: Today is the day of the.. Today is the day that before Korea holiday 설날, because, so, I went, I came to my hometown.

\par\medskip
\noindent
Example output:\\
\{\\
\hspace*{1em}"translated\_english\_phrase": ["Lunar New Year's Day", "Seollal"],\\
\hspace*{1em}"definition": "A traditional Korean holiday marking the first day of the lunar new year."\\
\}
\end{tcolorbox}

\section{\dataset stats}
\label{app:lake_stat}

The following plots show the distributions of sessions per learner, user utterances per session, learner-level CSW rates, and CSW function labels across two partners.

\begin{figure}[!ht]
    \centering
    \includegraphics[width=0.7\columnwidth]{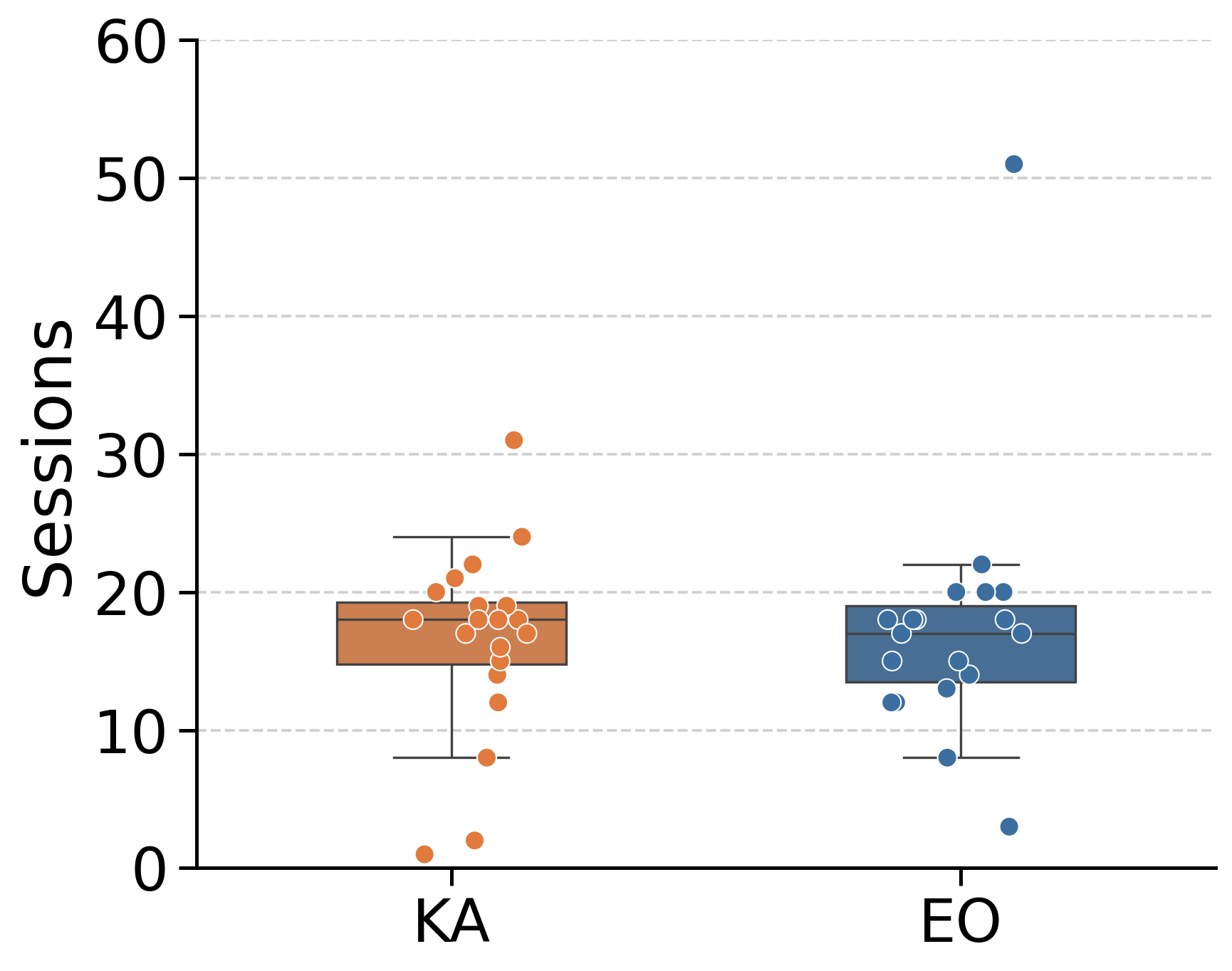}
    \caption{Distribution of session counts per learner.}
    \label{fig:lake_session_per_learner}
\end{figure}

\begin{figure}[!ht]
    \centering
    \includegraphics[width=0.8\columnwidth]{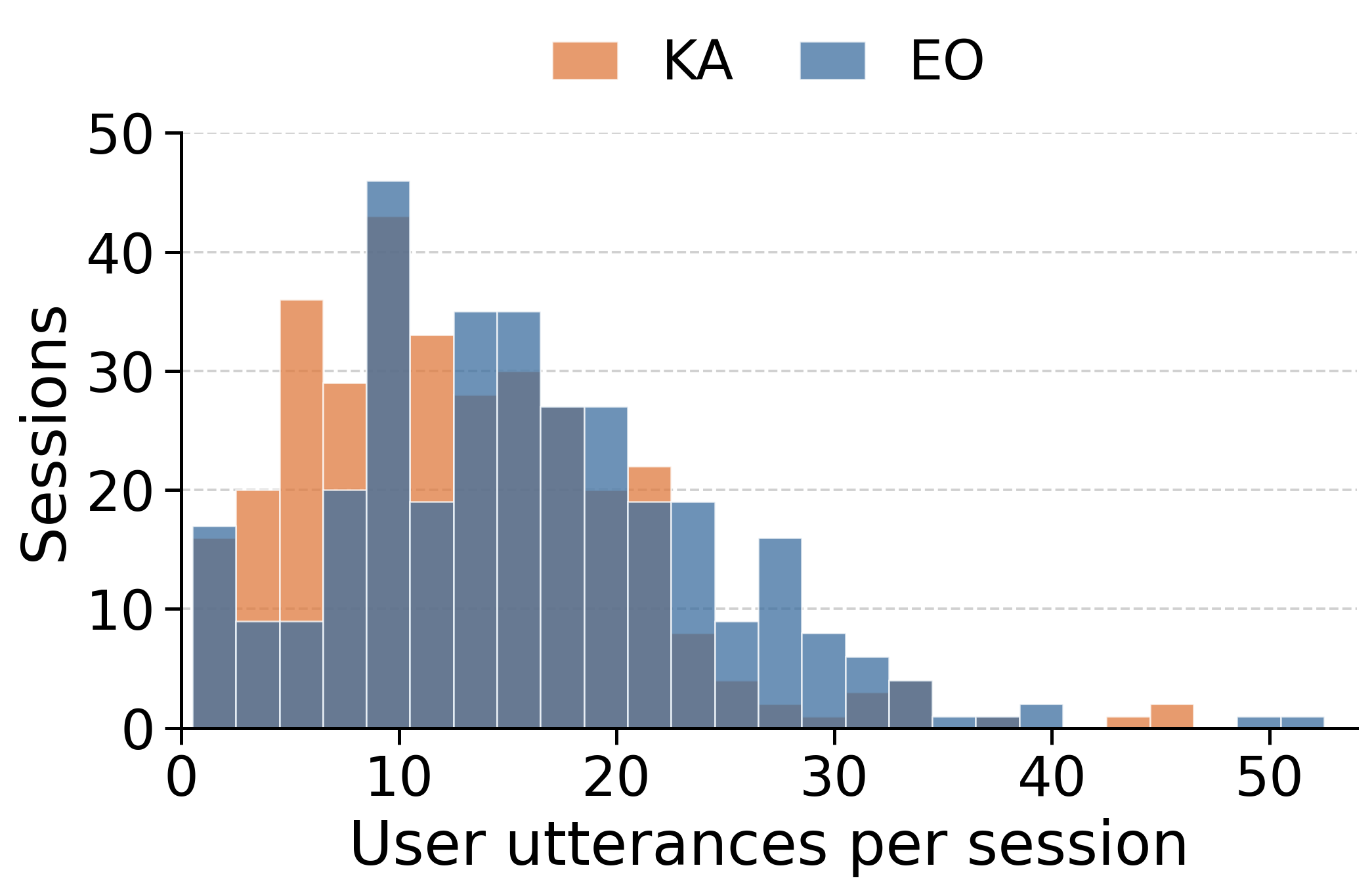}
    \caption{Distribution of session lengths (user utterances per session).}
    \label{fig:lake_session_length}
\end{figure}

\begin{figure}[!ht]
    \centering
    \includegraphics[width=0.8\columnwidth]{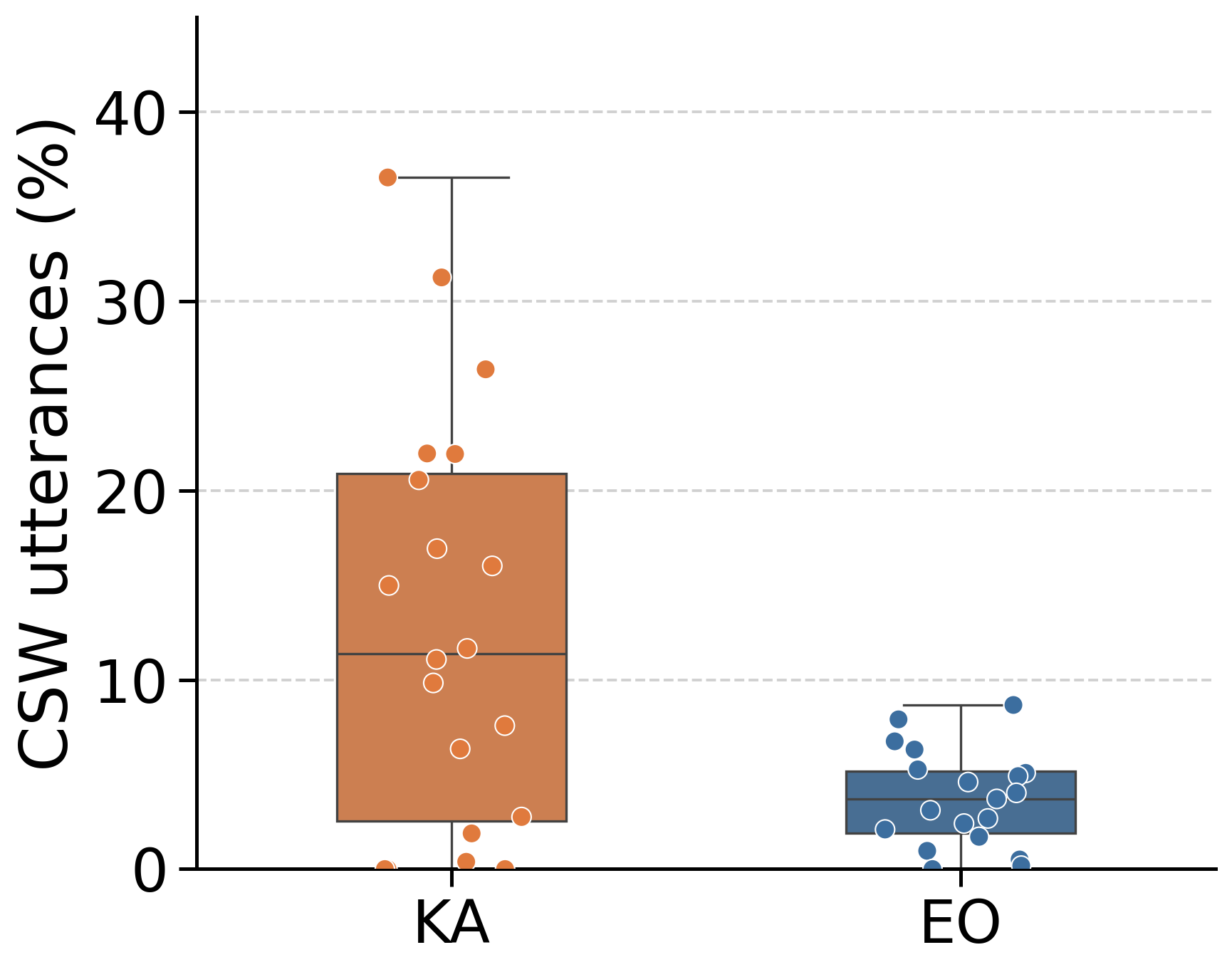}
    \caption{Distribution of CSW utterance percentage.}
    \label{fig:lake_csw_percentage}
\end{figure}

\begin{figure}[!ht]
    \centering
    \includegraphics[width=0.9\columnwidth]{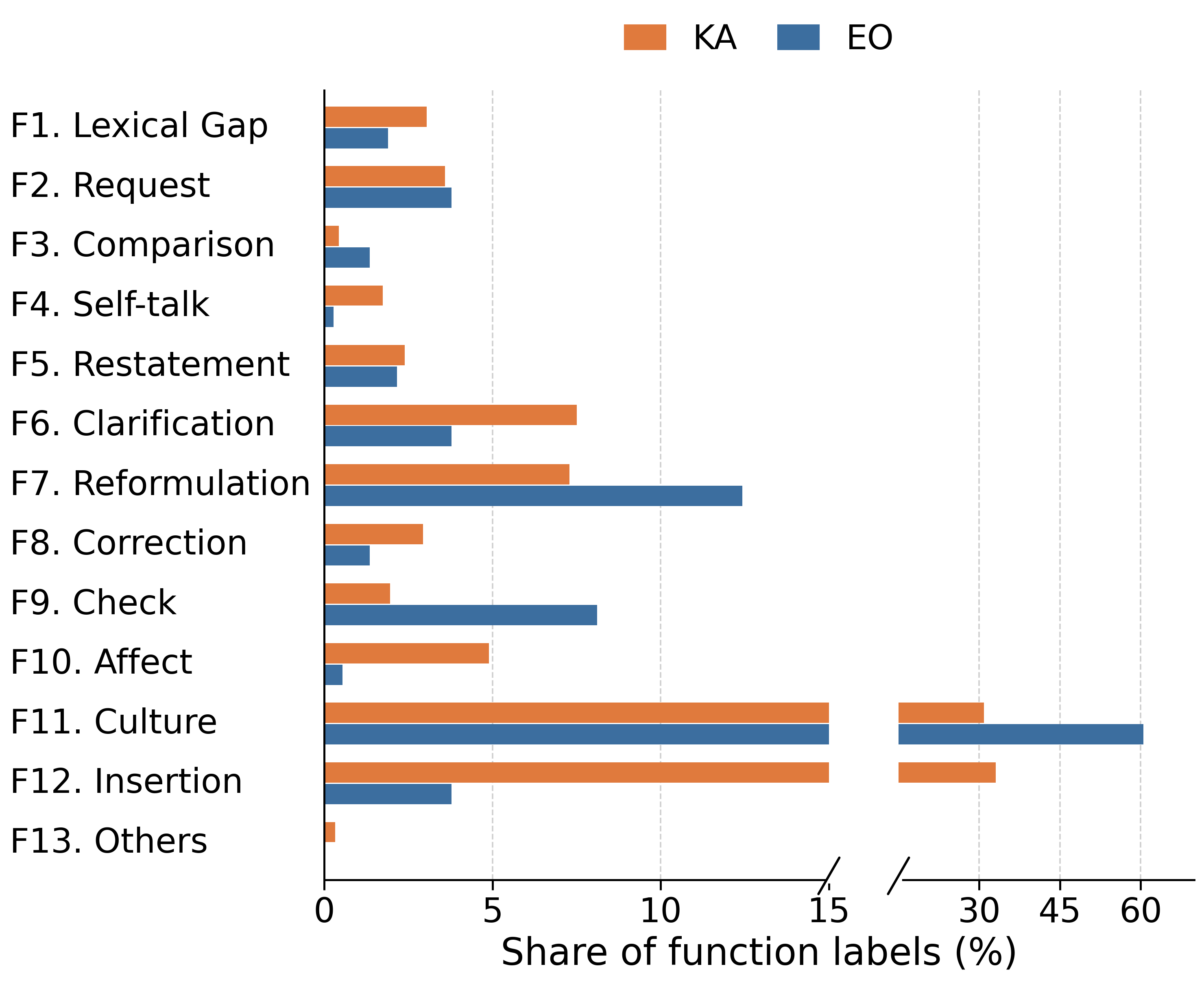}
    \caption{Distribution of CSW function labels.}
    \label{fig:lake_csw_functions}
\end{figure}

\section{CSW Function Taxonomy}
\label{app:csw_taxonomy_detail}

The final taxonomy comprises six categories and 13 functions (Table~\ref{tab:csw_taxonomy}).

\paragraph{Metalinguistic.}
This category includes switches that shift attention from conversational content to the learner's language production, building on prior accounts of metalinguistic activity and vocabulary search~\cite{yoon2023code, moore2002code, swain2000task}.
We distinguish an explicit lexical gap signal (\texttt{F1}) from an L2 equivalent request (\texttt{F2}), because signaling difficulty does not necessarily constitute an explicit request for assistance.
L1 and L2 comparison (\texttt{F3}) captures questions about the correspondence between an L1 expression and an L2 candidate.
Self-directed talk (\texttt{F4}) captures learners' comments to themselves while searching for an L2 expression, consistent with accounts of L1 use for self-regulation~\cite{anton1998socio}.
These distinctions preserve differences in how learners manage language production and what assistance they explicitly seek from the AI partner.

\paragraph{Meaning management.}
This category includes switches that draw on both languages to convey or clarify the learner's intended meaning, building on prior accounts of message clarification and bilingual meaning construction~\cite{yoon2023code, moore2002code, kim2012functions}.
We distinguish cross-language restatement (\texttt{F5}), clarification of L2 meaning through L1 (\texttt{F6}), and reformulation of an L1 expression in L2 (\texttt{F7}).
These distinctions capture how learners use the two languages to construct meaning, including the direction of clarification and reformulation.

\paragraph{Interactional.}
This category includes switches that address the AI partner's knowledge or interpretation of the learner's message.
Interlocutor knowledge check (\texttt{F9}) captures learners' curiosity about the partner's familiarity with L1 concepts, particularly those related to Korean culture.
In our dataset, interlocutor correction (\texttt{F8}) addresses misunderstandings that mainly arise from ASR errors or limitations in the AI partner's linguistic and cultural knowledge.
Although these uses are not commonly distinguished as separate functions in prior learner CSW taxonomies, they can be understood more broadly as establishing shared understanding and engaging in repair or meaning negotiation~\cite{moore2002code, noh2015study}.

\paragraph{Content.}
This category includes switches used to convey emotional meanings or culturally specific concepts as part of the conversational content.
Affective expression (\texttt{F10}) connects to established accounts of emotional expression through CSW~\cite{begum-etal-2016-functions, belani-flanigan-2023-automatic}.
Cultural reference (\texttt{F11}) builds on findings that Korean EFL learners use CSW to convey culture-specific meanings, including when they know an L2 equivalent~\cite{kim2012functions, yoon2023code}.
We group these functions to capture the affective and cultural significance of the content conveyed through L1.
Recognizing this significance can inform responses that acknowledge learners' emotions or engage with cultural meanings alongside support for L2 expression.

\paragraph{Unmarked.}
This category includes L1 insertions without an explicit signal indicating the purpose of the switch (\texttt{F12}).
Although such insertions may compensate for lexical difficulty, learners can also switch when they know an L2 equivalent~\cite{kim2012functions}.
An insertion alone therefore does not establish a lexical gap or a request for assistance.
We retain this category to record the absence of explicit functional cues without assigning an unsupported motivation.

\paragraph{Other.}
This category includes infrequent uses not captured by the remaining function labels (\texttt{F13}), such as quotation, an established CSW function~\cite{begum-etal-2016-functions, yoon2023code}.
We group these uses under a residual label because they occur rarely in our dataset.
Unlike \texttt{F12}, which records the absence of an explicit functional signal, \texttt{F13} captures identifiable uses outside the main function inventory.

\section{Learner CSW Function Classification}
\label{app:csw_function_classification}

\subsection{Experiment Setup}

We evaluate following models for CSW function classification accuracy:
\texttt{GPT-6 Luna}~\cite{openai2026gpt6luna},
\texttt{GPT-6.1 Sol}~\cite{openai2026gpt61},
\texttt{GPT-6 Astra}~\cite{openai2026gpt6astra},
\texttt{Gemini 3.8 Flash}~\cite{deepmind2026gemini38flash},
\texttt{Gemini 3.1 Pro}~\cite{deepmind2026gemini31pro},
and \texttt{Claude Sonnet 5.5}~\cite{anthropic2026claudesonnet55}

\begin{tcolorbox}[
    breakable,
    top=1pt,
    left=1pt,
    right=1pt,
    bottom=1pt,
    colback=white,
    fontupper=\ttfamily,
    fonttitle=\bfseries,
    title={Classification Prompt -- Utterance}
]
\small
Classify the code-switching function(s) of the following utterance. The target code-switched phrase is marked with << >>.\\

An utterance can have multiple functions. Return ONLY a valid JSON object with the key "functions" containing a list of the applicable function codes (e.g., ["F1", "F5"]). Do not include any explanation or additional text.

Taxonomy of Code-Switching Functions:\\

Metalinguistic\\
* F1: Lexical gap signal (e.g., "I can't remember the English word for <<종>>.")\\
* F2: L2 equivalent request (e.g., "What do you say <<사춘기>> in English?")\\
* F3: L1 and L2 comparison (e.g., "Is <<역할 놀이>> 'role playing' in English?")\\
* F4: Self-directed talk (e.g., "taste was... <<뭐라 해야 해>>... not so strong.")\\

Meaning Management\\
* F5: Meaning restatement in L1/L2 (e.g., "the <<목적>>, goal of the club")\\
* F6: L2 meaning clarification in L1 (e.g., "some bad smell, <<비린내>>")\\
* F7: L1 expression reformulation in L2 (e.g., "<<한 바퀴>>, one cycle, one circle")\\

Interactional\\
* F8: Interlocutor correction (e.g., "<<게장>> is crab. Not dog.")\\
* F9: Interlocutor knowledge check (e.g., "Do you know about <<설날>>?")\\

Content\\
* F10: Affective expression (e.g., "I feel so <<섭섭하다>>.")\\
* F11: Cultural reference (e.g., "eat <<떡국>> and do <<제사>>.")\\

Unmarked\\
* F12: L1 insertion without explicit signal (e.g., "I'm trying to write a <<자기소개서>>.")\\

Others\\
* F13: Others (e.g., quotation: "a part that says '<<우리 같이 저물자>>'")\\

Utterance: \{utterance\}
\end{tcolorbox}

\begin{tcolorbox}[
    breakable,
    top=1pt,
    left=1pt,
    right=1pt,
    bottom=1pt,
    colback=white,
    fontupper=\ttfamily,
    fonttitle=\bfseries,
    title={Classification Prompt -- Dialogue}
]
\small
Classify the code-switching function(s) of the following utterance. The target code-switched phrase is marked with << >>. The input is the whole dialogue up to and including that utterance. Only the target code-switched span is marked with << >>.\\

Taxonomy of Code-Switching Functions:\\

Metalinguistic\\
* F1: Lexical gap signal (e.g., "I can't remember the English word for <<종>>.")\\
* F2: L2 equivalent request (e.g., "What do you say <<사춘기>> in English?")\\
* F3: L1 and L2 comparison (e.g., "Is <<역할 놀이>> 'role playing' in English?")\\
* F4: Self-directed talk (e.g., "taste was... <<뭐라 해야 해>>... not so strong.")\\

Meaning Management\\
* F5: Meaning restatement in L1/L2 (e.g., "the <<목적>>, goal of the club")\\
* F6: L2 meaning clarification in L1 (e.g., "some bad smell, <<비린내>>")\\
* F7: L1 expression reformulation in L2 (e.g., "<<한 바퀴>>, one cycle, one circle")\\

Interactional\\
* F8: Interlocutor correction (e.g., "<<게장>> is crab. Not dog.")\\
* F9: Interlocutor knowledge check (e.g., "Do you know about <<설날>>?")\\

Content\\
* F10: Affective expression (e.g., "I feel so <<섭섭하다>>.")\\
* F11: Cultural reference (e.g., "eat <<떡국>> and do <<제사>>.")\\

Unmarked\\
* F12: L1 insertion without explicit signal (e.g., "I'm trying to write a <<자기소개서>>.")\\

Others\\
* F13: Others (e.g., quotation: "a part that says '<<우리 같이 저물자>>'")\\

Utterance: \{utterance\}
\end{tcolorbox}

\subsection{Experiment Results}
\begin{table}[h]
\centering
\small
\setlength{\tabcolsep}{5pt}
\begin{tabular}{lrr}
\toprule
\textbf{Model} & \textbf{Utterance} & \textbf{Dialogue} \\
\midrule
GPT-6 Luna        & 38.21 & 40.32 \\
GPT-6.1 Sol       & 71.08 & \textbf{72.30} \\
GPT-6 Astra       & \textbf{71.60} & 72.04 \\
Gemini 3.8 Flash  & 59.60 & 49.43 \\
Gemini 3.1 Pro    & 69.33 & 64.94 \\
Claude Sonnet 5.5 & 35.76 & 29.18 \\
\bottomrule
\end{tabular}
\caption{Automatic CSW function classification performance}
\label{tab:model_performance}
\end{table}
Treating each distinct, unordered label set as a single nominal category, we measure the exact-match agreement rate on the set of function labels. \texttt{GPT-6.1 Sol} shows the best classification performance with dialogue provided as context, achieving 72.3\%.

\section{Learner Simulation Prompts}
\label{app:simulation}

\begin{tcolorbox}[
    breakable,
    top=1pt,
    left=1pt,
    right=1pt,
    bottom=1pt,
    colback=white,
    fontupper=\ttfamily,
    fonttitle=\bfseries,
    title={CSW Learner Prompt (KA)}
]
\small
You are a South Korean undergraduate student learning English at an upper-intermediate (CEFR B2) level. Your native language is Korean. You are practicing your English conversation skills with an AI tutor who understands both English and Korean in a real-time spoken conversation. You naturally code-switch between English and Korean as a South Korean learner might in a real spoken conversation, but you do not have to code-switch in every response. Respond only as the learner. Generate output that reads like a raw spoken transcript, using English and Hangul as appropriate. Do not include explanations, translations in parentheses, brackets, stage directions, asterisks, or emojis.
\end{tcolorbox}

\begin{tcolorbox}[
    breakable,
    top=1pt,
    left=1pt,
    right=1pt,
    bottom=1pt,
    colback=white,
    fontupper=\ttfamily,
    fonttitle=\bfseries,
    title={CSW Learner Prompt (KA, Taxonomy)}
]
\small
You are a South Korean undergraduate student learning English at an upper-intermediate (CEFR B2) level. Your native language is Korean. You are practicing your English conversation skills with an AI tutor who understands both English and Korean in a real-time spoken conversation. You naturally code-switch between English and Korean as a South Korean learner might in a real spoken conversation, but you do not have to code-switch in every response. Respond only as the learner. Generate output that reads like a raw spoken transcript, using English and Hangul as appropriate. Do not include explanations, translations in parentheses, brackets, stage directions, asterisks, or emojis.\\

The following taxonomy describes code-switching functions identified among South Korean learners of English. You may refer to these functions when code-switching. The taxonomy is background context, not a checklist to reproduce. Do not include function codes or << >> markers in your responses. The brackets in the examples only mark the code-switched spans.\\

\{Taxonomy and Examples from Table~\ref{tab:csw_taxonomy}\}
\end{tcolorbox}

\begin{tcolorbox}[
    breakable,
    top=1pt,
    left=1pt,
    right=1pt,
    bottom=1pt,
    colback=white,
    fontupper=\ttfamily,
    fonttitle=\bfseries,
    title={CSW Learner Prompt (EO)}
]
\small
You are a South Korean undergraduate student learning English at an upper-intermediate (CEFR B2) level. Your native language is Korean. You are practicing your English conversation skills with an AI tutor who understands only English in a real-time spoken conversation. You naturally code-switch between English and Korean as a South Korean learner might in a real spoken conversation, but you do not have to code-switch in every response. Respond only as the learner. Generate output that reads like a raw spoken transcript, using English and Hangul as appropriate. Do not include explanations, translations in parentheses, brackets, stage directions, asterisks, or emojis.
\end{tcolorbox}

\begin{tcolorbox}[
    breakable,
    top=1pt,
    left=1pt,
    right=1pt,
    bottom=1pt,
    colback=white,
    fontupper=\ttfamily,
    fonttitle=\bfseries,
    title={CSW Learner Prompt (EO, Taxonomy)}
]
\small
You are a South Korean undergraduate student learning English at an upper-intermediate (CEFR B2) level. Your native language is Korean. You are practicing your English conversation skills with an AI tutor who understands only English in a real-time spoken conversation. You naturally code-switch between English and Korean as a South Korean learner might in a real spoken conversation, but you do not have to code-switch in every response. Respond only as the learner. Generate output that reads like a raw spoken transcript, using English and Hangul as appropriate. Do not include explanations, translations in parentheses, brackets, stage directions, asterisks, or emojis.\\

The following taxonomy describes code-switching functions identified among South Korean learners of English. You may refer to these functions when code-switching. The taxonomy is background context, not a checklist to reproduce. Do not include function codes or << >> markers in your responses. The brackets in the examples only mark the code-switched spans.\\

\{Taxonomy and Examples from Table~\ref{tab:csw_taxonomy}\}
\end{tcolorbox}

\section{Model Inference}
\paragraph{Proprietary models.}
We use APIs and inference infrastructure from OpenAI~\footnote{\url{https://openai.com/}}, Google AI Studio~\footnote{\url{https://aistudio.google.com/}}, and OpenRouter~\footnote{\url{https://openrouter.ai/}}.

\begin{itemize}
    \item \textbf{OpenAI.} We use the following models: \texttt{GPT-4o}, \texttt{GPT-4o Transcribe}, \texttt{GPT-4o Mini Transcribe}, \texttt{GPT-6 Luna}, \texttt{GPT-6.1-Sol}, \texttt{GPT-6 Astra}.
    \item \textbf{{Google AI Studio.}} We use the following models: \texttt{Gemini-3.8-Flash}, \texttt{Gemini 3.1 Pro}, \texttt{Gemini-3.5-Transcribe}.
    \item \textbf{{OpenRouter.}} We use the following model: \texttt{Claude-Sonnet-5.5}.
\end{itemize}

\paragraph{Open models.}
We use servers with one 80GB H100 GPU to run
\texttt{Gemma-4-12B-IT}, \texttt{Qwen3-ASR-1.7B} and \texttt{Qwen3-ASR-0.6B}, as well as six \texttt{Whisper} variants: \texttt{large-v3}, \texttt{large-v2}, \texttt{medium}, \texttt{small}, \texttt{base}, and \texttt{tiny}.

\end{document}